\documentclass[aps, 12pt]{revtex4}
\usepackage{graphicx,subfigure}
\usepackage{epsfig}

\usepackage{bbold}
\begin{document}
\title{Random Matrix theory approach to Quantum mechanics }
\author{K. V. S. Shiv Chaitanya}
\email[]{ chaitanya@hyderabad.bits-pilani.ac.in}
\affiliation{Department of Physics,, BITS Pilani, Hyderabad Campus, Jawahar Nagar, \\Shamirpet Mandal,
Hyderabad, India 500 078.}

\begin{abstract}
In this paper, we give random matrix theory approach to the quantum mechanics using the quantum 
Hamilton-Jacobi formalism. We show that the bound state problems in quantum mechanics are analogous to solving Gaussian unitary ensemble of random matrix theory. This study helps in identify the potential appear in the joint probability distribution function in the random matrix theory as a super potential. This approach allows to extend the random matrix theory to the newly discovered exceptional polynomials.
\end{abstract}

\maketitle

\section{Introduction}

Nine different  formulations of non-relativistic quantum mechanics exist  \cite{nine}. These are  the wavefunction,
matrix, path integral, phase space, density matrix, second quantization, variational, pilot wave, and
Quantum Hamilton-Jacobi (QHJ) formulations. 
It is interesting to note the close resemblance between random matrix theory and the QHJ formalism. The $n$ stationary points of the random matrix ensemble action \cite{met} $H=V(x)-\frac{1}{2}\sum_{k\neq l}log(\vert x_k-x_l\vert)$ where $V(x)$ is the potential. In QHJ formalism, the problem is solved using the quantum momentum function (QMF) and it plays an important role .  The quantum momentum function  is defined  as \cite{bhal,bhal1}  $p=\sum_{k=1}^n\frac{-i}{x-x_k}+Q(x)$ here the moving poles are simple poles with residue $-i\hbar$ (we take here $\hbar=m=1$) and $Q(x)$ is the singular part of the quantum momentum function.  The moving poles are described as  in the   $n$ first-order poles in the QMF of quantum mechanical origin. The positions of the  poles are and same as the zeros of the wave function. As the zeros of the
wave function change their positions with energy, so do the
location of the corresponding poles in the QMF  \cite{bhal}.  The moving poles behave like random numbers as they are energy dependent (as the energy changes the position of these moving poles will change). The above statement clearly captures the connection between the  $n$ stationary random points and the $n$ moving poles of QMF. 

There are other models in the literature whose equations resemble the random matrix ensemble action. They are the Stieltjes electrostatic problem \cite{st,st1} with $n$ moving unit charges, 
interacting  through a logarithmic potential, are placed
between  two fixed charges $p$ and $q$ at $a$ and $b$ respectively on a real line
is given by $V(x_k)+\sum_{k\neq l}\frac{1}{x_k-x_l}=0$ where $V(x_k)$ is
the position of the two fixed charges. The motion of point vortices in hydrodynamics is governed by the Helmholtz’s equations\cite{aref} $\frac{dz_k}{dt}=\frac{1}{2\pi i}\sum_{j=1}^{n}\frac{\Gamma_j}{z_k-z_j}+W(z_k)$ here the $\Gamma_j$ is circulations or strengths of point vortices at $z_j$ respectively and $W(z_k)$ is the back ground flow. The equation for low-energy spectrum of the Heisenberg ferromagnetic chain \cite{sas} $\frac{1}{u_l}=\pi d+\frac{2d}{l}\sum_{m\neq l}\frac{1}{u_l-u_m}$.  The analogy between the Stieltjes electrostatic problem  and the quantum momentum function is established by the author him self \cite{kvs} and the connection between the Heisenberg ferromagnetic chain and 
the Stieltjes electrostatic problem was established in ref \cite{sas}. 

In this paper, we show  that the solving quantum mechanics using QHJ formalism is analogous to solving the quantum mechanics problem using the random matrix theory. The paper is organised as follows, in the following section of the introduction we briefly describe the RMT (part A) followed by QHJ (part B) and Supersymmetric quantum mechanics (part C). In the section II, we give the equivalence between the QHJ and RMT followed by conclusion in section III.

\subsection{Random Matrix Theory} 

In 1951, Wigner proposed the use of random matrix theory (RMT) to describe the properties of excited states of atomic nuclei \cite{wig}. This
was the first time RMT was used to model physical reality. In random matrix theory, the dynamics of the ensemble of a random infinite dimensional Hermitian matrix is described by the probability distribution function 
\begin{equation}\label{pdfo}
{\cal P}(\lambda_1,\ldots,\lambda_N)\,d\Lambda
= c_ne^{-\beta H}
\,\,d\Lambda
\end{equation}
where  $d\Lambda=d\lambda_1 \ldots d\lambda_N$, $c_n$ is constant of proportionality and the index $\beta=1, 2, 4$  characterizes the real parameters of the symmetry class of orthogonal, unitary and symplectic respectively for the random  matrix 
\begin{equation}
H=-\sum_{i=1}^NV(\lambda_i)-\sum_{i<1}^Nln\vert\lambda_i-\lambda_j\vert.
\end{equation}
 As the random matrix is invariant under the symmetry group these ensembles are called the invariant ensembles. In the random matrix $H$,  $V(\lambda_i)$ is the potential, $\lambda_i$ are the eigenvalues with $i$ being a free index running from $1,2,\cdots N$ and $\vert\lambda_i-\lambda_j\vert$ is the Vandermonde determinant. For details please refer to \cite{met}. The probability distribution function can be rewritten as 
\begin{equation}\label{pdf}
{\cal P}(\lambda_1,\ldots,\lambda_N)\,d\Lambda
=c_n e^{-\sum_{i=1}^N\beta V(\lambda_i)}
\prod_{i<j}\vert\lambda_i-\lambda_j\vert^\beta
\,\,d\Lambda
\end{equation}
By manipulating the Vandermonde determinant, that is, by adding and deleting columns or rows, the equation (\ref{pdf}) is written as
\begin{equation}\label{pdf1}
{\cal P}(\lambda_1,\ldots,\lambda_N)\,d\Lambda
=c_n \prod_{i=1}^N w^{\frac{1}{2}} (\lambda_i)
\prod_{i<j}\vert\lambda_i-\lambda_j\vert
\,\,d\Lambda
\end{equation}
where $w_\beta (\lambda_i)$ is the weight function of classical orthogonal polynomials.
The classical orthogonal polynomials are classified into three different 
categories depending upon the range of the polynomials. The polynomials in the intervals $(-\infty;\infty)$ with weight function $w=e^{-\lambda^2}$ are the Hermite polynomials.
In the intervals $[0;\infty)$ with weight function $w=\lambda^be^{-\lambda}$ are the Laguerre polynomials. In the intervals $[-1;1]$ with weight function $w=(1+\lambda)^{a}(1-\lambda)^b$ are the Jacobi polynomials.

The invariant ensembles in random matrix theory are classified into three the Gaussian ensembles, the Wishart ensembles and the two Wishart ensembles.  In the case of Gaussian, the matrices are known as the Wigner matrices. The Gaussian ensembles are further classified the  Gaussian orthogonal ensemble (GOE), the Gaussian
unitary ensemble (GUE), and the Gaussian symplectic ensemble (GSE).  In stationary case, for the Gaussian ensembles, Wigner law states that the empirical distribution of eigenvalues $\lambda_1,\ldots,\lambda_N$ as $N$ tends to infinity converges to a semicircle in the interval [a,b]. In non-stationary case the equation (\ref{pdf}) is the stationary solution of a Fokker-Planck equation \cite{met}. 

In this paper we only deal with the Gaussian ensembles and show that the bound state problems in quantum mechanics are analogous to solving Gaussian unitary ensemble of random matrix theory. This study helps in identifying the potential that appear in the joint probability distribution function in the random matrix theory as a super potential. This approach allows to extend the random matrix theory to the newly discovered exceptional polynomials.

\subsection{Quantum Hamilton Jacobi}

In this section, a brief review of  Quantum Hamilton Jacobi formalism is presented below.
For details see the references \cite{sree}. 
The Schr\"odinger equation is given by,
\begin{equation}
- \frac{\hbar^2}{2m}\nabla^2\psi(x,y,z)+ V(x,y,z) \psi(x,y,z) = E
  \psi(x,y,z).    \label{sc} 
\end{equation}
One defines a function $S$ analogous to the classical characteristic function by the relation 
\begin{equation}
\psi(x,y,z) = \exp\left(\frac{iS}{\hbar}\right)       \label{ac}
\end{equation}
which, when substituted in (\ref{sc}), gives
\begin{equation}
(\vec{\nabla}S)^2 -i \hbar \vec{\nabla}.(\vec{\nabla}S) = 2m (E
  - V(x,y,z)).   \label{qhj0} 
\end{equation}
the quantum momentum function $p$ is defined in terms of the function $S$ as
\begin{equation}
\vec{p} = \vec{\nabla} S. \label{mp}
\end{equation}
Substituting (\ref{mp}) in (\ref{qhj0}) gives the QHJ equation
for $\vec{p} $ as 
\begin{equation}
(\vec{p})^2 - i \hbar \vec{\nabla}.\vec{p} = 2m (E - V(x,y,z))
  \label{bhy} 
\end{equation}
and from (\ref{sc}) and (\ref{mp}), one can see that $\vec{p}$ 
is the the logarithmic derivative of $\psi(x,y,z)$ {\it i. e}. 
\begin{equation}
\vec{p} = -i \hbar \vec{\nabla} ln \psi(x,y,z) \label{lg}
\end{equation}
The above discussion of the QHJ formalism is done in three
dimensions the same equation in one dimension takes the following form 
\begin{equation}
p^2 - i \hbar \frac{dp}{dx} = 2m (E - V(x)),       \label{qhj1}
\end{equation}
which is also known as the Riccati equation. In one dimension the eq (\ref{lg}) take the form
\begin{equation}
p = -i \hbar \frac{d}{dx}ln \psi(x). \label{lg1}
\end{equation}
It is shown by Leacock and Padgett \cite{qhj, qhj1} that the action angle variable 
gives rise to exact quantization condition
\begin{equation}
J(E) \equiv  \frac{1}{2\pi} \oint_C{pdx} = n\hbar.       \label{act}
\end{equation}
\subsection{SUSY Quantum mechanics}
In supersymmetry, the superpotential $\mathcal{W}(x)$ is defined in terms of the intertwining operators $ \hat{A}$ and $\hat{A}^{\dagger}$  as
\begin{equation}
  \hat{A} = \frac{d}{dx} + \mathcal{W}(x), \qquad \hat{A}^{\dagger} = - \frac{d}{dx} + \mathcal{W}(x), 
\label{eq:A}
\end{equation}
This allows one to define a pair of factorized Hamiltonians $H^{\pm}$ as
\begin{eqnarray}
   H^{+} &=& 	\hat{A}^{\dagger} \hat{A} 	= - \frac{d^2}{dx^2} + \mathcal{V}^{+}(x) - E, \label{vp}\\
  H^{-} &=& 	\hat{A}  {\hat A}^{\dagger} 	= - \frac{d^2}{dx^2} + \mathcal{V}^{-}(x) - E, \label{vm}
\end{eqnarray}
where $E$ is the factorization energy. 
The partner potentials $\mathcal{V}^{\pm}(x)$ are related to $\mathcal{W}(x)$ by 
\begin{equation}\label{gh}
 \mathcal{V}^{\pm}(x) = \mathcal{W}^2(x) \mp \mathcal{W}'(x) + E, 
\end{equation}
where  prime denotes differentiation with respect to $x$,refer the reader to \cite{kharebook} for more details. 

\section{Equivalence between RMT and QHJ}
To establish the relation between the RMT and QHJ we start with the quantum momentum function given in equation (\ref{qhj1}). In equation (\ref{qhj1}) $p$ is the QMF and is given by equation (\ref{lg1}). It is evident  form the equation  (\ref{lg1}) the QMF is related to wave function. Further more the relation between the QMF and the wave function in the equation (\ref{lg1}) captures the randomness arising from the moving poles in QMF. 
Then the quantum momentum function in terms of $n$ moving poles and fixed poles \cite{bhal, bhal1, sree, geo1} is given by 
\begin{equation}
p(x)=\sum_{k=1}^n\frac{-i\hbar}{x-x_k}+Q(x).\label{uf}
\end{equation}
By introducing the polynomial
\begin{eqnarray}
f(x)=(x-x_1)(x-x_2)\cdots (x-x_n),\label{poly}
\end{eqnarray}
then the quantum momentum function ( $\hbar=1$) in terms of the polynomial reads as 
\begin{equation}
p=\sum_{k=1}^n i\frac{f'(x)}{f(x)}+Q(x)\label{uf4}
\end{equation}
and substituting in (\ref{qhj1}) then one gets
\begin{eqnarray}
f''(x) + 2iQ(x)f'(x)+[Q^2(x)-iQ'(x)-E+V(x)]f(x)=0.\label{dif11}
\end{eqnarray} 
The search for the polynomial solutions to the equation (\ref{dif11}) leads to quantization. This is equivalent to demanding $ [Q^2(x)-iQ'(x)-E + V(x)]$ to be constant. This will only be possible  if 
$Q(x)=i\mathcal{W}(x)$.  In QHJ it turns out that the $Q(x)$ is the super potential.  It should be pointed out here that the energy in the equation (\ref{dif11}) id different from the energy in equation (\ref{gh}). Then the equation (\ref{dif11}) reduces to
\begin{eqnarray}
-f''(x) +2\mathcal{W}(x)f'(x)=-E_nf(x).\label{dif111}
\end{eqnarray} 
It is evident from the above equation that the left hand side ( \textit{lhs}) balances the right hand side ( \textit{rhs}).
By dividing equation (\ref{dif111}) with $2f'(x_k)$ and defining the \textit{lhs} of equation (\ref{dif111}) as T(x) one gets
\begin{eqnarray}
T(x)=-\frac{f''(x_k)}{2_n'(x_k)} +\mathcal{W}(x).\label{di1}
\end{eqnarray} 

Now, we will show that QHJ is analogous to  the Steiljes electrostatic model  and then show that QHJ is  equivalent to RMT. 
By following the argument in the Metha's book \cite{met}
\begin{eqnarray}
\sum_{k=1}\frac{1}{x-x_k}=\frac{f'(x)}{f(x)}-\frac{1}{x-x_j}=\frac{(x-x_j)f'(x)-f(x)}{(x-x_j)f(x)},\label{dig}
\end{eqnarray}
by taking the following limit $x\rightarrow x_j$ and using l'Hospital rule one gets
\begin{eqnarray}
\sum_{1\leq j\leq n,j\neq k} \frac{1}{x_j-x_k}&=&\lim_{x\rightarrow x_j}
\left[\frac{f'(x)}{f(x)}-\frac{1}{x-x_j}\right]\nonumber\\&=&\lim_{x\rightarrow x_j}\frac{(x-x_j)f'(x)-f(x)}{(x-x_j)f(x)}
\nonumber \\&=& \frac{f''(x_j)}{2f'(x_j)}.\label{lhp}
\end{eqnarray}
Substituting (\ref{lhp}) in (\ref{di1}) one gets
\begin{equation}
T(x)=\sum_{1\leq j\leq n,j\neq k} \frac{1}{x_k-x_j} + \mathcal{W}(x),\label{ufo}
\end{equation}
which is minimum of the Lagrangian considered for the the Steiljes electrostatic model \cite{st,st1}.

By re-substituting $Q(x)=i\mathcal{W}(x)$
in equation (\ref{ufo})  the quantum momentum function reduces to
\begin{equation}
p(x_j)=\sum_{1\leq j\leq n,j\neq k} \frac{-i}{x_k-x_j}+Q(x_j).\label{uf1}\;\; k=1,2\cdots n.
\end{equation}

In order to compare the RMT and QHJ, we define the probability distribution function in RMT as the wave function 
\begin{equation}
\psi(\lambda_k)={\cal P}(\lambda_1,\ldots,\lambda_N)=C_n exp^{-\beta H}\label{rwf}
\end{equation}
where 
\begin{equation}
H=V(\lambda_k)-\frac{1}{2}\sum_{1\leq j\leq n,j\neq k}ln(\vert \lambda_k-\lambda_j\vert)
\end{equation} 
Substituting (\ref{rwf}) in (\ref{lg1}) the quantum momentum function is given by
\begin{eqnarray}\label{uf11}
p(\lambda_k)=\sum_{1\leq j\leq n,j\neq k}\frac{\beta}{\lambda_k-\lambda_j}-S(\lambda_k)\;\; k=1,2\cdots n
\end{eqnarray}
where $S(x_k)=\beta\frac{dV(\lambda_l)}{d\lambda_k}$. By comparing equation (\ref{uf11}) and (\ref{uf}) one gets $\beta=-i \hbar$ and 
$\mathcal{W}(x)=\frac{1}{\beta}S(x)=Q(x)$. It is therefore clear that the potential used in the random matrix theory is the super potential $\mathcal{W}(x)$.
Substituting the equation (\ref{poly}) in  equation (\ref{uf11}) and the substuting this in equation (\ref{qhj1}) one gets the following differential equation  as
\begin{eqnarray}
f''(x) + 2iQ(x)f'(x)+[Q^2(x)-iQ'(x)-E+V(x)]f(x)=0.\label{fgy}
\end{eqnarray} 

By exploting the supersymmetric quantum mechanics one gets the solutions for the above differential equation as the classical orthogonal polynomials for the given range fixed by the potential. 
 For the Harmonic oscillator \cite{kharebook} the super potential is given by  $Q(x)=\frac{1}{2}\omega x$ then the differential equation (\ref{fgy}) becomes
\begin{eqnarray}
f''(\zeta) - 2\zeta f'(x)+2nf(\zeta)=0 \label{herm} ~~~~1\leq k\leq n.
\end{eqnarray} 
where $\zeta=\alpha x$ and $\alpha^2= \frac{\omega}{2}$. The differential equation (\ref{herm}) is the Hermite differential equation and the solutions are the Hermite polynomials given by $\psi(x)= exp[-\frac{1}{2}x^2] H_n(x)$.  One can show by supersymmetric arguments the potential is given by $V(x)=\frac{1}{2}\omega x^2$. This model in RMT is known as the Dyson gas model.

One can extend the RMT models to the recently discovered exceptional polynomials, as quantum Hamiltonian Jacobi formalism for the deformed oscillator is studied in ref \cite{pkp} and the momentum function is given by  
\begin{eqnarray}\label{hj}
p=\sum_{k=1}^{n}\frac{1}{x-x_k}-x-\frac{1}{x-x_1}+\frac{g+l}{x}.
\end{eqnarray}
and thus the solutions are exceptional Laguerre polynomials 
we obtain the solution as exceptional polynomials
\begin{eqnarray}
\psi=\frac{x^{g+l}e^{-\frac{1}{2}x^2}}{L^{g+l - \frac{3}{2}}_l(−x^2)}\hat{L}^n_l(x^2,g).
\end{eqnarray}

It is clear from quantum mechanic the wave function is given by 
$\psi = w^{\frac{1}{2}}(x)P_l(x)$
where $w(x)$ is the weight function for the corresponding orthogonal polynomial $P_l$. The probability of finding the particle in an interval is given by 
\begin{eqnarray}
\rho=\int_a^b \psi(x)^*\psi(x)dx=\int_a^b w(x)P_l(x)P_m(x)dx.
\end{eqnarray}
In random matrix theory this turns out to be the Joint probability distribution function
Gaussian unitary ensemble in terms of eigenvalues 
\begin{eqnarray}\label{pdf23}
{\cal P}(\lambda_1,\ldots,\lambda_N)\,d\Lambda
=C_n w\vert\Delta(x)\vert^2\,\,d\Lambda.
\end{eqnarray} 
where $ \vert\Delta(x)\vert =\prod_{i<j}\vert\lambda_i-\lambda_j\vert$ is the Vandermonde determinant.  This the Joint probability distribution function for exceptional Laguerre polynomials in terms of eigenvalues is given by
\begin{eqnarray}\label{pdf11}
{\cal P}(\lambda_1,\ldots,\lambda_N)\,d\Lambda
=C_n \frac{x^{2(g+l)}e^{- x^2}}{L^{2(g+l)-3}_l(−x^2)}
\prod_{i<j}\vert\lambda_i-\lambda_j\vert^2
\,\,d\Lambda.
\end{eqnarray}  
where $\hat{L}^n_l(x^2,g)$ are exceptional Laguerre polynomials.

\section{Conclusion}

In this paper, we given random matrix theory approach to the quantum mechanics using the quantum 
Hamilton-Jacobi formalism. We shown that the bound state problems in quantum mechanics are analogous to solving Gaussian unitary ensemble of random matrix theory. We have also shown that the  potential appear in the joint probability distribution function in the random matrix theory as a super potential. Using this approach we have extend the random matrix theory to the newly discovered exceptional polynomials.

\section{Acknowledgements}
KVSSC acknowledges the Department of Science and technology, Govt of India, (fast track
scheme (D. O. No: SR/FTP/PS-139/2012)) for financial support. Author thank  A. K. Kapoor, V. Srinivasan, Kannan Ramaswamy and P K Thirivikraman for stimulating conversations. 

%%%%%%%%%%%%%%%%%%%%%%%%
%%%%%%%%%%%%%%%%%%%%%%%%

\end{document}